\documentstyle[aps,psfig,graphics,multicol]{revtex}
\begin{document}
\draft

\title{
Ising model in scale-free networks: A Monte Carlo simulation}
\author{Carlos P. Herrero}
\address{Instituto de Ciencia de Materiales,
         Consejo Superior de Investigaciones Cient\'{\i}ficas (CSIC),
         Campus de Cantoblanco, 28049 Madrid, Spain  } 
\date{\today}
\maketitle

\begin{abstract}
The Ising model in uncorrelated scale-free networks has been studied by means
of Monte Carlo simulations. These networks are characterized by a degree
(or connectivity) distribution $P(k) \sim k^{-\gamma}$. 
The ferromagnetic-paramagnetic transition temperature 
has been studied as a function of the parameter $\gamma$. 
For $\gamma > 3$ our results agree with earlier analytical calculations, which
found a phase transition at a temperature $T_c(\gamma)$ in the thermodynamic limit.
For $\gamma \leq 3$,  a ferromagnetic-paramagnetic crossover occurs at a
size-dependent temperature $T_{co}$, and the system is in the ordered ferromagnetic
state at any temperature for a system size $N \to \infty$.
For $\gamma = 3$ and large enough $N$, the crossover temperature is found to be
$T_{co} \approx A \ln N$, with a prefactor $A$ proportional to the mean degree.
For $2 < \gamma < 3$, we obtain $T_{co} \sim \langle k \rangle N^z$, with an 
exponent $z$ that decreases as $\gamma$ increases. This exponent is found to be 
lower than predicted by earlier calculations. 
 \\
\end{abstract}
%

\begin{multicols}{2}
         
Complex networks describe several kinds of natural and artificial systems
(social, biological, technological, economic), and are currently employed
as models to study various processes taking place in real-life systems 
\cite{st01,al02a,do02a}.
In last years, new models of complex networks have been
introduced, motivated by empirical data in different fields. 
Thus, the so-called small-world \cite{wa98} and scale-free (SF) networks \cite{ba99}
incorporate various aspects of real systems.
These complex networks provide us with the underlying topological structure to study
processes such as spread of infections \cite{mo00}, signal propagation \cite{st01,he02},
and cooperative phenomena \cite{ba00,le02,go03,do03}. 

In a SF network the degree distribution, $P(k)$, where $k$ is the number
of links connected to a node, has a power-law decay $P(k) \sim k^{-\gamma}$.
This kind of networks have been found in particular in social systems \cite{ne01},
in protein interaction networks \cite{je01}, in the internet \cite{si03},
and in the world-wide web \cite{al99}.
In both natural and artificial networks, the exponent $\gamma$ controlling the
degree distribution is usually in the range $2 < \gamma < 3$ \cite{do02a,go02}. 

Cooperative phenomena in complex networks are expected to display unusual
characteristics, associated to the peculiar topology of these systems.
In this context, the Ising model on SF networks has been studied with several
theoretical techniques \cite{le02,do02b,ig02}, and its critical behavior 
was found to be dependent on the exponent $\gamma$. In particular,
when $\langle k^2 \rangle$ is finite, there appears a ferromagnetic (FM) to 
paramagnetic (PM) transition at a finite temperature $T_c$.  
On the contrary, when $\langle k^2
\rangle$ diverges (as happens for $\gamma \leq 3$), the system remains in its 
ordered FM phase at any temperature, and no phase transition occurs in the
thermodynamic limit. 

Here we investigate the FM-PM transition for the Ising model in scale-free
networks with various values of the exponent $\gamma$.
We employ Monte Carlo (MC) simulations to obtain the transition temperature, and
compare it with that predicted in earlier calculations.  Our results confirm
those of analytical calculations for $\gamma > 3$, and are used to check the
precision of those obtained earlier with approximate methods for $\gamma \leq 3$. 

Our networks are defined, apart from $\gamma$, by the maximum and
minimum degrees, denoted $k_{\rm cut}$ and $k_0$, respectively. 
Thus, the number of sites with degree $k$ is given by
$N_k = (k_{\rm cut} / k)^{\gamma}$ for $k_0 \leq k \leq k_{\rm cut}$, and
$N_k = 0$ otherwise.
This gives $N_{k_{\rm cut}} = 1$, and a system size $N = \sum_{k_0}^{k_{\rm cut}}
N_k$, which for large $k_{\rm cut}$ scales as $N \sim k_{\rm cut}^{\gamma}$.
Once $\{N_k\}$ is defined [or the corresponding probability density $P(k) = N_k/N$], 
one has a total number of ends of links (total degree) 
$K = \sum_{k_0}^{k_{\rm cut}} k N_k$.
Then we ascribe a degree to each node according to $\{N_k\}$, and finally connect at 
random ends of links (giving a total of $L = K/2$ connections), with the conditions: 
(i) no two nodes can have more than one bond connecting them, and 
(ii) no node can be connected by a link to itself.
We have checked that networks generated in this way are uncorrelated, in the
sense that the joint probability $P(k,k')$ for degrees of nearest neighbors
fulfills the relation 
$P(k,k') = k \, k' P(k) P(k') / \langle k \rangle^2$ \cite{do02a}.

On these scale-free networks, we consider the Hamiltonian
\begin{equation}
H = - \sum_{i < j} J_{ij} S_i S_j   \, ,
\end{equation}
where $J_{ij} = J (> 0)$ if nodes $i$ and $j$ are connected, and $J_{ij} = 0$ 
otherwise. $S_i = \pm 1$ ($i = 1, ..., N$) are Ising spin variables.
Sampling of the configuration space has been carried out by the Metropolis
local update algorithm \cite{bi97}. This allows us to study the temperature 
dependence of the magnetization, and in particular the transition from a FM to a 
PM regime as $T$ is increased.  Depending on the value of the exponent
$\gamma$, this transition: (i) can occur at a well-defined temperature in the
thermodynamic limit ($N \to \infty$), or (ii) may display a FM-PM crossover 
temperature shifting with system size and diverging to infinity as
$N \to \infty$.
For the sake of clarity we will employ a different notation for the temperature
of phase change in both cases. In the first case, we will call it $T_c$ as for 
thermodynamic phase transitions ($T_c<\infty$). In the second case, it will be called
crossover temperature, $T_{co}$, to emphasize its size dependence. 

\begin{figure}
\vspace*{-1.7cm}
\centerline{\psfig{figure=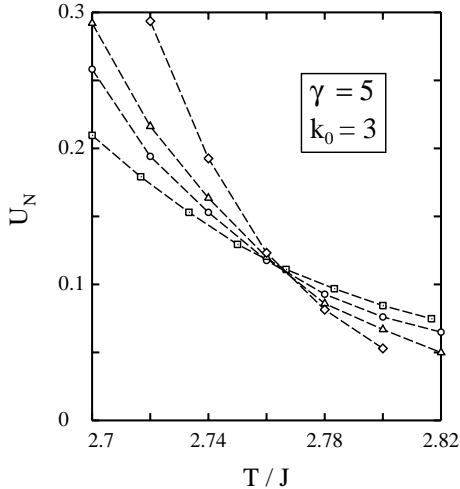,height=10.0cm}}
\vspace*{-1.5cm}
\caption{
Fourth-order Binder's cumulant $U_N$ as a function of temperature for
scale-free
networks with $\gamma = 5$ and $k_0 = 3$.  Symbols represent different system
sizes:
squares, $k_{\rm cut}$ = 15; circles, $k_{\rm cut}$ = 17; triangles, $k_{\rm
cut}$ = 19,
and diamonds, $k_{\rm cut}$ = 22. These values of $k_{\rm cut}$ correspond to
system sizes $N$ ranging from 4300 to 29300.
} \label{f1} \end{figure}

When a FM-PM transition occurs in the thermodynamic limit, 
the transition temperature $T_c$ has been determined by using Binder's fourth-order 
cumulant\cite{bi97}
\begin{equation} \label{Binder}
U_N(T) \equiv 1 - \frac{ \langle M^4 \rangle_N } {3 \langle M^2 \rangle^2_N} \, ,
\end{equation} 
where the magnetization $M$ of a given spin configuration $\{S_i\}$ is 
$M = \sum_{i=1}^N S_i / N$. The average values in Eq.\,(\ref{Binder}) are taken
over different network realizations and different spin configurations for
a given network at temperature $T$.  In this case, the transition temperature
is obtained from the unique crossing point for several sizes $N$. 

In the second case, the size-dependent crossover temperature $T_{co}(N)$ has 
been determined from the maximum of the magnetization fluctuations, 
$(\Delta M)^2_N = \langle M^2 \rangle_N - \langle M \rangle^2_N$, as a function of 
temperature. We have checked that the crossover temperatures obtained by using
this criterion agree within error bars with those derived from the maximum derivative 
of the heat capacity. In particular, both procedures give the same size-dependence
of $T_{co}$ in the cases presented below.  

The largest networks considered here included about $5 \times 10^5$ sites.  
Such network sizes are required in particular to determine the power law
characterizing the size dependence of the crossover temperature $T_{co}$ for 
scale-free networks with $2 < \gamma < 3$. On the contrary, for the cases in
which a phase transition exists in the thermodynamic limit ($\gamma > 3$),
smaller sizes are necessary (see below).
The results presented below were obtained by averaging in each case over 1000
networks, except for the largest system sizes, for which 400 network
realizations were considered.

{\it Case $\gamma > 3$.}---
For scale-free networks with $\gamma > 3$, the average value $\langle k^2 \rangle$
converges to a finite value as $k_{\rm cut} \to \infty$. In this case, analytical
calculations \cite{le02,do02b} predict a well-defined FM-PM transition temperature $T_c$ 
given by
\begin{equation} \label{Tcanal}
\frac{J}{T_c} = \frac{1}{2}  \ln \left( \frac {\langle k^2 \rangle}
        { \langle k^2 \rangle - 2 \langle k \rangle } \right)  \; .
\end{equation} 
We have calculated Binder's cumulant $U_N$ for several values of $\gamma$ and 
different network sizes.
As an example, in Fig. 1 we present $U_N$ as a function of temperature for $\gamma = 5$,
$k_0 = 3$, and various values of $k_{\rm cut}$.  The transition temperature
is obtained from the crossing point for different system sizes 
(or cutoffs $k_{\rm cut}$).

The same procedure has been repeated for other values of $k_0$. The resulting values
of $T_c$ are presented in Fig. 2 (open symbols), along with the transition temperature
predicted by Eq.\,(\ref{Tcanal}) (solid line). 
In fact, this line was obtained by joining values of $T_c$ derived from
Eq.\,(\ref{Tcanal}) for integer values of $k_0$ in the limit $k_{\rm cut} \to \infty$.
The MC results agree within error bars with the transition temperature given by
Eq.\,(\ref{Tcanal}).
For comparison, we also present the critical temperature obtained in a simple mean-field
approach \cite{le02}: $T_c^{MF} = \langle k^2 \rangle / \langle k \rangle$, which is 
displayed as a dashed line.

\begin{figure}
\vspace*{-1.7cm}
\centerline{\psfig{figure=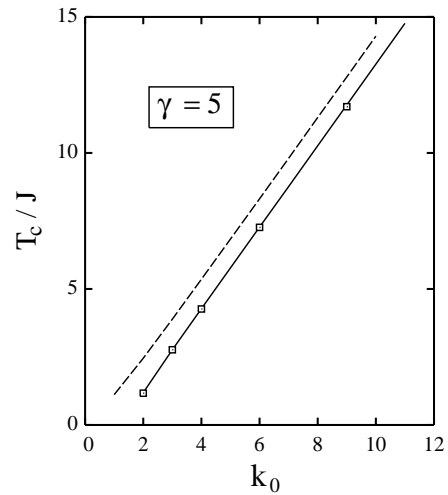,height=10.0cm}}
\vspace*{-1.5cm}
\caption{
Transition temperature $T_c$ for scale-free networks with $\gamma = 5$ as a
function of the minimum degree $k_0$.
Symbols represent results of MC simulations, as obtained from Binder's
cumulant.
Error bars are less than the symbol size.
The solid line was plotted by connecting points obtained from
Eq.\,(\ref{Tcanal}) for
integer values of $k_0$ and $k_{\rm cut} \to \infty$.
The dashed line shows the mean-field result for $T_c$.
} \label{f2} \end{figure}

The critical temperature $T_c$ obtained from Eq.\,(\ref{Tcanal}) for $\gamma = 5$
and $k_0 > 3$ can be fitted
linearly with good precision as $T_c = a k_0 + b$, with the parameters
$a = 1.50$ and $b = -1.72$.  The value of $a$ can in fact be estimated by 
approximating by integrals the sums giving the average values $\langle k \rangle$
and $\langle k^2 \rangle$ in Eq.\,(\ref{Tcanal}). This approach gives for large $k_0$, 
$T_c \approx \frac{3}{2} k_0$.
We note that for $k_0$ = 2 and 3, the actual values of $T_c$ deviate slightly from such a 
linear fit.
For $k_0$ = 1, Eq.\,(\ref{Tcanal}) is not defined, since 
$\langle k^2 \rangle - 2 \langle k \rangle < 0$. In this case, the
simulated networks consist of many different (not connected) components, and Binder's 
cumulant does not give a crossing for different system sizes.
 
{\it Case $\gamma = 3$.}---
The transition temperature given by Eq.\,(\ref{Tcanal}) increases for increasing $\gamma$ 
and eventually diverges for $\gamma \to 3$, as a consequence of the divergence of 
$\langle k^2 \rangle$.
For $\gamma = 3$, analytical calculations \cite{do02b,ig02} predict a FM-PM
crossover at a size-dependent temperature $T_{co}$, which scales as $\log N$. 
Such a logarithmic increase of $T_{co}$ with $N$ has been also obtained by 
Aleksiejuk {\em et al.} \cite{al02b,al02c} from Monte Carlo simulations of the 
Ising model 
in Barab\'asi-Albert growing networks. We note that these networks have $\gamma = 3$, but
display correlations between degrees of adjacent nodes \cite{ba99}.

\begin{figure}
\vspace*{-1.7cm}
\centerline{\psfig{figure=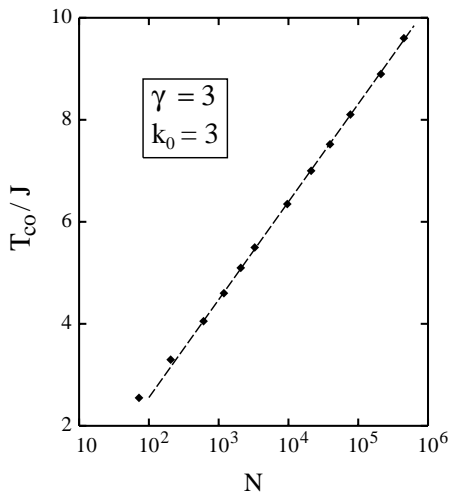,height=10.0cm}}
\vspace*{-1.5cm}
\caption{
Crossover temperature $T_{co}$ for scale-free networks with $\gamma = 3$ and $k_0
= 3$,
as a function of the system size $N$, presented in a logarithmic plot.
Symbols indicate results derived from MC simulations, with error bars less than
the
symbol size. The dashed line is a least-square fit to the data points for $N >
500$.
} \label{f3} \end{figure}

Our results for $T_{co}$ in the case $\gamma = 3$ and $k_0 = 3$ are shown in Fig. 3 as a 
function of the system size $N$ in a logarithmic plot. We indeed find a logarithmic 
dependence of $T_{co}$ on $N$, as in earlier works. In Fig. 3, symbols 
represent simulation
results and the dashed line is a least-square fit to the data points with $N > 500$ 
(smaller sizes give $T_{co}$ values that deviate from the asymptotic trend).
Thus, our results indicate a dependence: $T_c/J = A \ln N + B$, with constants
$A = 0.83$ and $B = -1.28$.
We have repeated the MC simulations for $k_0$ = 5 and 9, and obtained the same
logarithmic dependence as for $k_0 = 3$. The prefactor $A$ increases linearly with 
$k_0$, and in fact we found: $A/k_0 = 0.28 \pm 0.01$.
For Barab\'asi-Albert networks with $k_0 = 5$, Aleksiejuk {\em et al.}\cite{al02b}
found from a fit similar to ours $A = 2.6$, which means $A/k_0 = 0.52$.

For uncorrelated scale-free networks with $\gamma = 3$, Dorogovtsev {\em et al.}
\cite{do02b} found $T_{co}/J \approx \frac{1}{4} \langle k \rangle \ln N$. For 
$k_0 = 3$ and $k_{\rm cut} \to \infty$, one has $\langle k \rangle$ = 5.125, and 
thus their calculations predict $T_{co}/J \approx 1.28 \ln N$, with a prefactor 
$A$ on the order of unity, as that obtained here.
Mean-field calculations for $\gamma = 3$ give $T_c = \frac12 J k_0 \ln N$
\cite{le02,al02c}, which translates into a ratio $A/k_0 = 0.5$, somewhat larger 
than that found from our MC simulations.

\begin{figure}
\vspace*{-1.7cm}
\centerline{\psfig{figure=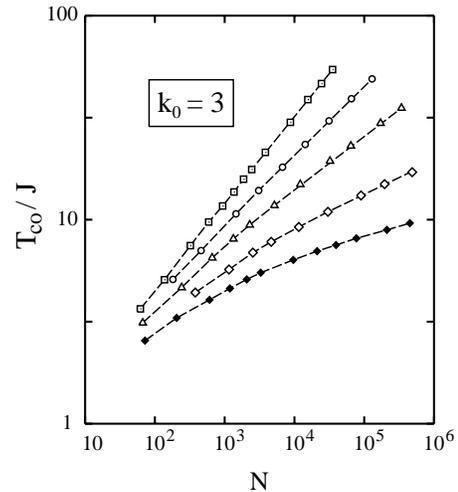,height=10.0cm}}
\vspace*{-1.5cm}
\caption{
Crossover temperature $T_{co}$ for scale-free networks with $2 \leq \gamma \le 3$,
as a function of the system size $N$, in a log-log plot. From top to bottom:
$\gamma$ = 2, 2.2, 2.4, 2.7, and 3. Error bars are less than the symbol size.
Dashed lines are guides to the eye.
} \label{f4} \end{figure}

{\it Case $2 < \gamma < 3$.}---
For scale-free networks with $\gamma < 3$, analytical calculations \cite{do02b,ig02} 
predict a size-dependent crossover temperature $T_{co}$ scaling as 
$\sim J \langle k \rangle N^z$, with an exponent $z$ dependent of the parameter $\gamma$.
In Fig. 4 we show the temperature $T_{co}$ as a function of the system size
$N$ for several values of $\gamma$ in a log-log plot, as derived from our MC
simulations (for $k_0 = 3$). 
The exponent $\gamma$ decreases from top to bottom: $\gamma$ = 2, 2.2, 2.4, 2.7, and 3.
For a given system size, $T_{co}$ decreases as $\gamma$ increases, as a consequence 
of the reduction in $\langle k^2 \rangle$.
 For a given $\gamma < 3$ and large-enough networks, $\log T_{co}$ displays a linear
dependence on $\log N$, as expected for a crossover temperature $T_{co}$ 
diverging as a power of the system size $N$. 
This linear dependence is obtained for system sizes $N \gtrsim N_0$, $N_0$
increasing with $\gamma$ and eventually diverging for $\gamma \to 3$. This means that
the present MC procedure cannot be applied to obtain accurately the exponent $z$ close to
$\gamma = 3$, unless one employes much larger system sizes. However, for $\gamma < 2.8$, 
$z$ can be found with enough precision for the system sizes considered here.

Thus, we have derived the exponent $z$ from our simulation results for $\gamma < 2.8$
by obtaining the slope of $\log[T_{co} / (J \langle k \rangle)]$ vs $\log N$ for
large $N$.  Our results are shown in Fig. 5 (open symbols) as a function of $\gamma$. 
The solid line represents the analytical prediction \cite{do02b,ig02}:
\begin{equation} \label{zz}
  z = \frac {3 - \gamma} {\gamma - 1} \;  .
\end{equation}
Our results agree with the analytical calculations in that the ratio
$T_{co}/(J \langle k \rangle)$
diverges with $N$ as a power law, with exponent $z$ increasing for decreasing
$\gamma$.  However, our MC simulations give values of $z$ lower than Eq.\,(\ref{zz}).
In particular, for $\gamma \to 2$ Eq.\,(\ref{zz}) gives $z = 1$, and our numerical
procedure yields $z = 0.43 \pm 0.02$.

\begin{figure}
\vspace*{-1.7cm}
\centerline{\psfig{figure=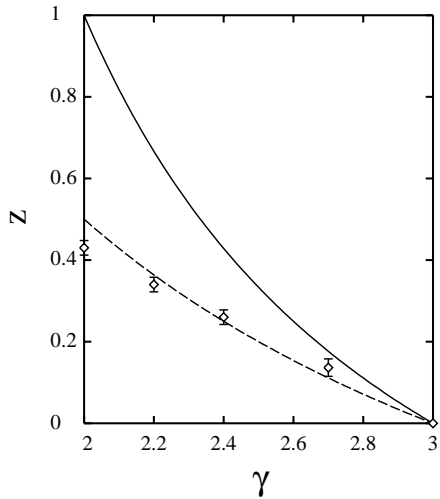,height=10.0cm}}
\vspace*{-1.5cm}
\caption{
The exponent $z$ giving the power-law dependence of the crossover temperature
$T_{co}$ is plotted vs the parameter $\gamma$ for scale-free networks.
Open symbols are results derived from Monte Carlo simulations.
The solid line corresponds to the analytical prediction given by Eq.\,(\ref{zz}).
The dashed line represents the dependence $z = (3 - \gamma) / \gamma$.
} \label{f5} \end{figure}

The exponent $z$ in Eq.\,(\ref{zz}) is related to the divergence of
$\langle k^2 \rangle$ in networks with $\int_{k_{\rm cut}}^{\infty} P(k) dk \sim 1/N $,
which means that $k_{\rm cut} \sim N^{1/(\gamma-1)}$ \cite{do02b,ig02}. 
However, for the networks considered here, for practical computational reasons we
have defined a sharp cutoff $k_{\rm cut}$, which gives a dependence
$\langle k^2 \rangle \sim N^{(3-\gamma)/\gamma}$, or $k_{\rm cut} \sim N^{1/\gamma}$.
Therefore, assuming that $T_{co}$ diverges with the same exponent as
$\langle k^2 \rangle$, we have $z = (3-\gamma)/\gamma$. This dependence of $z$ on
$\gamma$ is plotted in Fig. 5 as a dashed line. Our MC results follow this line,
but separate from it at $\gamma = 2$. It is not clear the reason for 
this discrepancy. One can argue, however, that in the limit $\gamma \to 2$ the 
average value $\langle k \rangle$ diverges logarithmically, and thus the convergence
of the exponent for the ratio $\langle k^2 \rangle / \langle k \rangle$ should be
very slow, in spite of the apparent convergence shown in Fig. 4.
This point may require further theoretical consideration.

In summary, we have studied the FM-PM transition for the Ising
model in uncorrelated scale-free networks, by means of Monte Carlo simulations.
For $\gamma > 3$ our results for the temperature transition fully agree with earlier
analytical calculations, confirming the appearance of a well-defined transition in
the thermodynamic limit. For $\gamma \leq 3$ we find a crossover temperature which
increases with system size. In particular, for $\gamma = 3$ such an increase
is found to be $T_{co} \approx 0.28 k_0 \ln N$, whereas for $\gamma < 3$ we obtained
$T_{co} \sim J \langle k \rangle N^z$, with an exponent $z$ lower than predicted
by earlier analytical calculations.  We finally note that some care should be taken 
when comparing analytical results for scale-free netwoks with those derived from
simulations, since the cutoff definition for the actually simulated networks 
may appreciably change the results in some cases.

The author benefited from useful discussions with E. Chac\'on and E. Velasco.
This work was supported by CICYT (Spain) under Contract 
No. BFM2000-1318.  \\

\end{multicols}
\end{document}